\documentclass[prb,aps, twocolumn, amsmath,amssymb,superscriptaddress]{revtex4}

\usepackage{bibunits}

\usepackage{float}
\usepackage{amsmath}

\usepackage{amssymb}
\usepackage{amsfonts}
\usepackage{euscript}
\usepackage{enumerate}
\usepackage{hhline}
\usepackage{pslatex}
\usepackage{tabularx}
\usepackage[usenames,dvipsnames]{xcolor}

\usepackage{graphicx}
\usepackage{dcolumn}
\usepackage{bm}
\usepackage[sort&compress]{natbib}

\makeatletter
\renewcommand{\@biblabel}[1]{#1. }
\renewcommand{\@dotsep}{500}
\renewcommand{\@pnumwidth}{0em}
\renewcommand{\l@figure}[2]{
\@dottedtocline{1}{1.5em}{2em}{Figure #1}{}\vspace{15pt}}

\begin{document}
\begin{bibunit}

\title{Chip-integrated visible-telecom photon pair sources for quantum communication}

\author{Xiyuan Lu}\email{xiyuan.lu@nist.gov}
\affiliation{Center for Nanoscale Science and Technology, National
Institute of Standards and Technology, Gaithersburg, MD 20899,
USA}\affiliation{Maryland NanoCenter, University of Maryland,
College Park, MD 20742, USA}
\author{Qing Li}
\affiliation{Center for Nanoscale Science and Technology, National
Institute of Standards and Technology, Gaithersburg, MD 20899,
USA}\affiliation{Maryland NanoCenter, University of Maryland,
College Park, MD 20742, USA}
\author{Daron A. Westly}
\affiliation{Center for Nanoscale Science and Technology, National
Institute of Standards and Technology, Gaithersburg, MD 20899, USA}
\author{Gregory Moille}
\affiliation{Center for Nanoscale Science and Technology, National
Institute of Standards and Technology, Gaithersburg, MD 20899,
USA}\affiliation{Maryland NanoCenter, University of Maryland,
College Park, MD 20742, USA}
\author{Anshuman Singh}
\affiliation{Center for Nanoscale Science and Technology, National
Institute of Standards and Technology, Gaithersburg, MD 20899,
USA}\affiliation{Maryland NanoCenter, University of Maryland,
College Park, MD 20742, USA}
\author{Vikas Anant}
\affiliation{Photon Spot, Inc., Monrovia, CA 91016}
\author{Kartik Srinivasan} \email{kartik.srinivasan@nist.gov}
\affiliation{Center for Nanoscale Science and Technology, National
Institute of Standards and Technology, Gaithersburg, MD 20899, USA}

\date{\today}

\begin{abstract}
\noindent \textbf{Photon pair sources are fundamental building blocks for quantum entanglement and quantum communication. Recent studies in silicon photonics have documented promising characteristics for photon pair sources within the telecommunications band, including sub-milliwatt optical pump power, high spectral brightness, and high photon purity. However, most quantum systems suitable for local operations, such as storage and computation, support optical transitions in the visible or short near-infrared bands. In comparison to telecommunications wavelengths, the significantly higher optical attenuation in silica at such wavelengths limits the length scale over which optical-fiber-based quantum communication between such local nodes can take place. One approach to connect such systems over fiber is through a photon pair source that can bridge the visible and telecom bands, but an appropriate source, which should produce narrow-band photon pairs with a high signal-to-noise ratio, has not yet been developed in an integrated platform.  Here, we demonstrate a nanophotonic visible-telecom photon pair source for the first time, using high quality factor silicon nitride resonators to generate bright photon pairs with an unprecedented coincidence-to-accidental ratio (CAR) up to ${(\rm 3800~\pm~200)}$.  We further demonstrate dispersion engineering of the microresonators to enable the connection of different species of trapped atoms/ions, defect centers, and quantum dots to the telecommunications bands for future quantum communication systems.}
\end{abstract}


\maketitle

Long-distance quantum communication requires resources for the low-loss storage and transmission of quantum information~\cite{nicolas_gisin_quantum_2007, lvovsky_optical_2009}. For fiber optic networks, this requirement presents a challenge, in that many systems that are best suited for storage, including trapped ions, neutral atoms, and spins in crystals, are connected to photons in the visible or short near-infrared (NIR) bands~\cite{simon_quantum_2010}, where losses in silica fibers are orders of magnitude higher than at telecommunications wavelengths~\cite{miya_ultimate_1979}.  While a quantum frequency conversion interface can bridge the spectral gap between the relevant wavelengths~\cite{raymer_manipulating_2012}, it can be quite challenging, both in finding appropriate laser sources to match the input and output wavelengths, and also in limiting added noise photons due to the strong classical pump fields.  These challenges are more acute as the local quantum system's operating wavelength decreases and moves further away from the telecommunications band.

\begin{center}
\begin{figure*}
\begin{center}
\includegraphics[width=\linewidth]{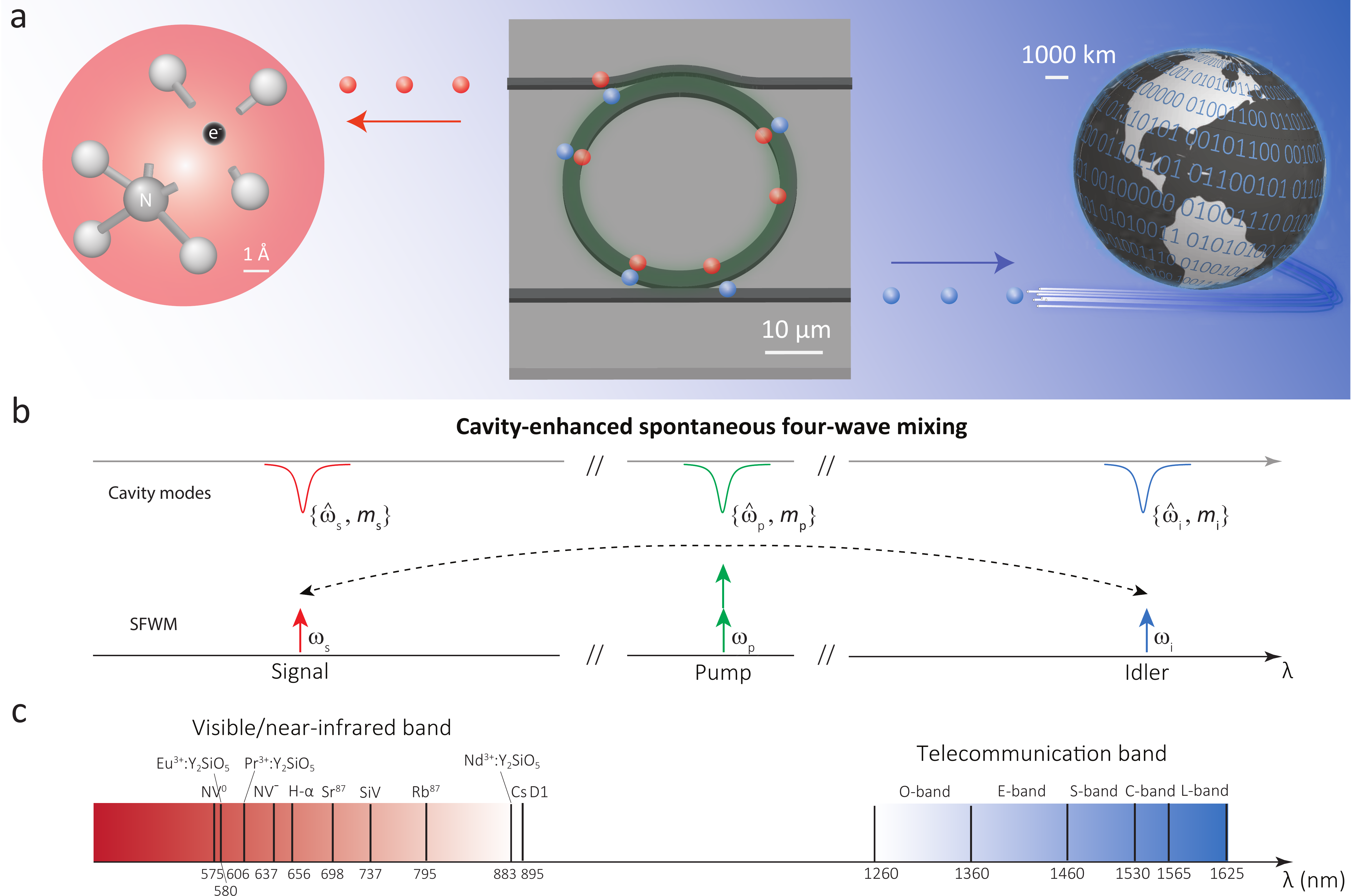}
\caption{\textbf{Motivation and Scheme}. \textbf{a}, Intended application, where photon pairs generated from a silicon nitride nanophotonic chip (center) connect the visible and telecommunication bands, to enable a link between a nitrogen vacancy center in diamond (left) and the telecom fiber network (right), for example. \textbf{b}, Scheme for cavity-enhanced spontaneous four-wave mixing (SFWM).  The continuous wave, classical pump field at frequency $\omega_{p}$ generates signal and idler photons at $\omega_{s}$ and $\omega_{i}$, respectively, dictated by energy conservation $2\omega_{p} = \omega_{s} + \omega_{i}$.  The nearest relevant cavity mode frequencies are at $\hat{\omega}_{p}$, $\hat{\omega}_{s}$, and $\hat{\omega}_{i}$, respectively. Phase-matching of light generated in these optical modes is equivalent to the condition on the cavity azimuthal mode numbers $m$ that $2m_{p} = m_{s} + m_{i}$.  \textbf{c}, Examples of local quantum systems with optical transitions in the visible/short near-infrared band (left), and classification of the telecommunications wavelength bands (right).}
\label{fig:Fig1}
\end{center}
\end{figure*}
\end{center}

A second approach to connecting photonic quantum systems over distance is through suitably-engineered entangled photon pair sources. In particular, quantum memories can be remotely entangled via entanglement swapping~\cite{pan_experimental_1998,halder_entangling_2007}, in which photons entangled with the two distant memories propagate towards each other and meet, where they are subjected to a Bell-state measurement. For visible/short-NIR memories, the propagation distance in optical fiber is limited as described above.  To overcome this limitation, one can introduce entangled photon pair sources in which one photon is resonant with the memory (i.e., at a visible wavelength), and the other is in the telecom band (Fig.~\ref{fig:Fig1}).  Successive entanglement swapping operations can then entangle the memories, with their separation now limited by the propagation distance in optical fiber in the telecom. Such visible-telecom pair sources might also find use in related scenarios, for example, Bell inequality violations using nitrogen vacancy center spins in diamond~\cite{hensen_loophole-free_2015}.

Existing visible-telecom photon pair sources are in platforms such as periodically-poled nonlinear crystals~\cite{clausen_source_2014}, photonic crystal fibers \cite{soller_bridging_2010}, and millimeter-scale crystalline microresonators~\cite{schunk_interfacing_2015}. The first two platforms generate broadband photons, so that spectral filtering to achieve the narrow linewidths needed for interaction with local quantum systems must be implemented. Filtering after pair generation comes at the price of a decrease in collection efficiency and signal-to-noise ratio, so incorporation of the nonlinear medium into a macroscopic cavity~\cite{fekete_ultranarrow-band_2013} has been pursued.  The crystalline microresonator platform, though promising, currently exhibits a coincidence-to-accidental ratio (CAR) below 20, which can be a constraint in applications. In general, these platforms all lack the potential for scalable fabrication and integration common to the planar systems used in nanophotonics.

A nanophotonic device platform for narrow-linewidth, high-quality visible-telecom photon pairs is the whispering gallery mode microresonator based on silicon photonics~\cite{caspani_integrated_2017}. Recent studies have documented promising characteristics for photon pair sources based on spontaneous four-wave mixing (SFWM) in the telecommunications band, including sub-milliwatt optical pump power, high spectral brightness, and high photon purity~\cite{jiang_silicon-chip_2015,xiyuan_lu_biphoton_2016,grassani_micrometer-scale_2015,savanier_photon_2016,ramelow_silicon-nitride_2015,jaramillo-villegas_persistent_2017},
and the potential for large-scale integration~\cite{wang_multidimensional_2018}. While silicon's narrow bandgap forbids operation below 1.1~$\mu$m, stoichiometric silicon nitride (Si$_3$N$_4$) has a much wider bandgap, enabling operation across wavelengths from the near-UV to the mid-infrared. Such broadband performance has been exploited in octave-spanning optical frequency combs~\cite{okawachi_octave-spanning_2011,li_stably_2017,karpov_photonic_2018}, and such microcombs have been operated sub-threshold to generate high-dimensional, frequency-bin entangled photon pairs in the telecom band~\cite{kues_-chip_2017,imany_50-ghz-spaced_2018}. Similar dispersion-engineered Si$_3$N$_4$ microresonators have been used to achieve efficient single-photon-level frequency conversion interfaces between the 980~nm and 1550~nm bands~\cite{li_efficient_2016}.  Here, we demonstrate a family of visible/short NIR-telecom photon pair sources using the Si$_3$N$_4$ microresonator platform.  In particular, we show that it is possible to overcome the primary challenge in the development of such sources, which is in achieving phase- and frequency-matching of narrow-band modes over a spectral separation of $>$250~THz.  Our source produces photons at 668~nm and 1548~nm, with a photon pair flux of ${\rm (7.0~\pm~0.8) \times 10^5}$ counts per second and a CAR of ${\rm (44~\pm~3)}$ at 0.46 mW input power. The CAR increases to ${\rm (3800~\pm~200)}$ at a pair flux of ${\rm (1.2~\pm~0.3) \times 10^3}$ counts per second. As discussed in the Supplementary Information (Fig. S1), this competes favorably with existing macroscopic-scale and mm-scale technologies, and is comparable to the best telecom-band sources based on silicon microresonators. Moreover, we experimentally show that the visible photons can be shifted from 630 nm to 810 nm by varying the pump wavelength and microring width. Our demonstration of a chip-integrated visible-telecom photon pair source emitting bright, pure, and narrow linewidth photons has the potential to serve as an important resource in connecting quantum memories with telecommunication networks.

\noindent \textbf{System Design}
The device is a Si$_3$N$_4$ microring resonator with a lower silicon dioxide (SiO$_2$) cladding and a top air cladding. Si$_3$N$_4$ microrings support high quality factor ($Q$) whispering-gallery modes, which combine a long photon lifetime with localization into a small effective volume ($V$). This leads to a large optical intensity (i.e., $I \sim Q/V$), which is critical for the SFWM efficiency \cite{boyd_nonlinear_2003,agrawal_nonlinear_2007}, that is, in reducing pump power and excess noise. Moreover, the broad transparency window of Si$_3$N$_4$/SiO$_2$, and the ability to compensate for material dispersion through waveguide dispersion, are critical to engineering visible-telecom photon pair sources. In this section, we outline the basic design procedure.

Unlike the linear momentum conservation criterion in a waveguide ($\Delta \beta = 0$, where $\beta$ is the propagation constant), phase-matching of angular momentum is required in a microring. As the electromagnetic field of a whispering gallery mode has an azimuthal dependence that scales as ${\rm{exp}}(im\phi)$, where $\phi$ is the azimuthal angle and $m$ is the mode number, phase-matching becomes a mode number matching criterion ($\Delta m = 0$). For degenerately-pumped spontaneous four-wave mixing (SFWM), the criterion is $2m_{p}-m_{s}-m_{i}=0$, where $m_{p}/m_{s}/m_{i}$ are the azimuthal mode numbers for the pump/signal/idler modes.  Next, assuming an appropriate set of modes is found to satisfy this requirement, the energy conservation condition imposed by SFWM ($2\omega_{p}-\omega_{s}-\omega_{i}=0$) must involve pump/signal/idler fields ($\omega_{p}/\omega_{s}/\omega_{i}$) that are resonant with the corresponding cavity modes ($\hat{\omega}_{p}/\hat{\omega}_{s}/\hat{\omega}_{i}$), with a maximum frequency mismatch dictated by the $Q$ of the corresponding cavity mode, e.g., $\delta\omega_{p,s,i} = |\omega_{p,s,i} - \hat{\omega}_{p,s,i}| \lesssim \hat{\omega}_{p,s,i}/2Q_{p,s,i}$. This challenge is one of matching frequencies separated over hundreds of THz with an accuracy within a cavity linewidth ($\approx$~1~GHz for a loaded $Q$ of $\rm 10^5$).

\begin{center}
\begin{figure*}
\begin{center}
\includegraphics[width=0.8\linewidth]{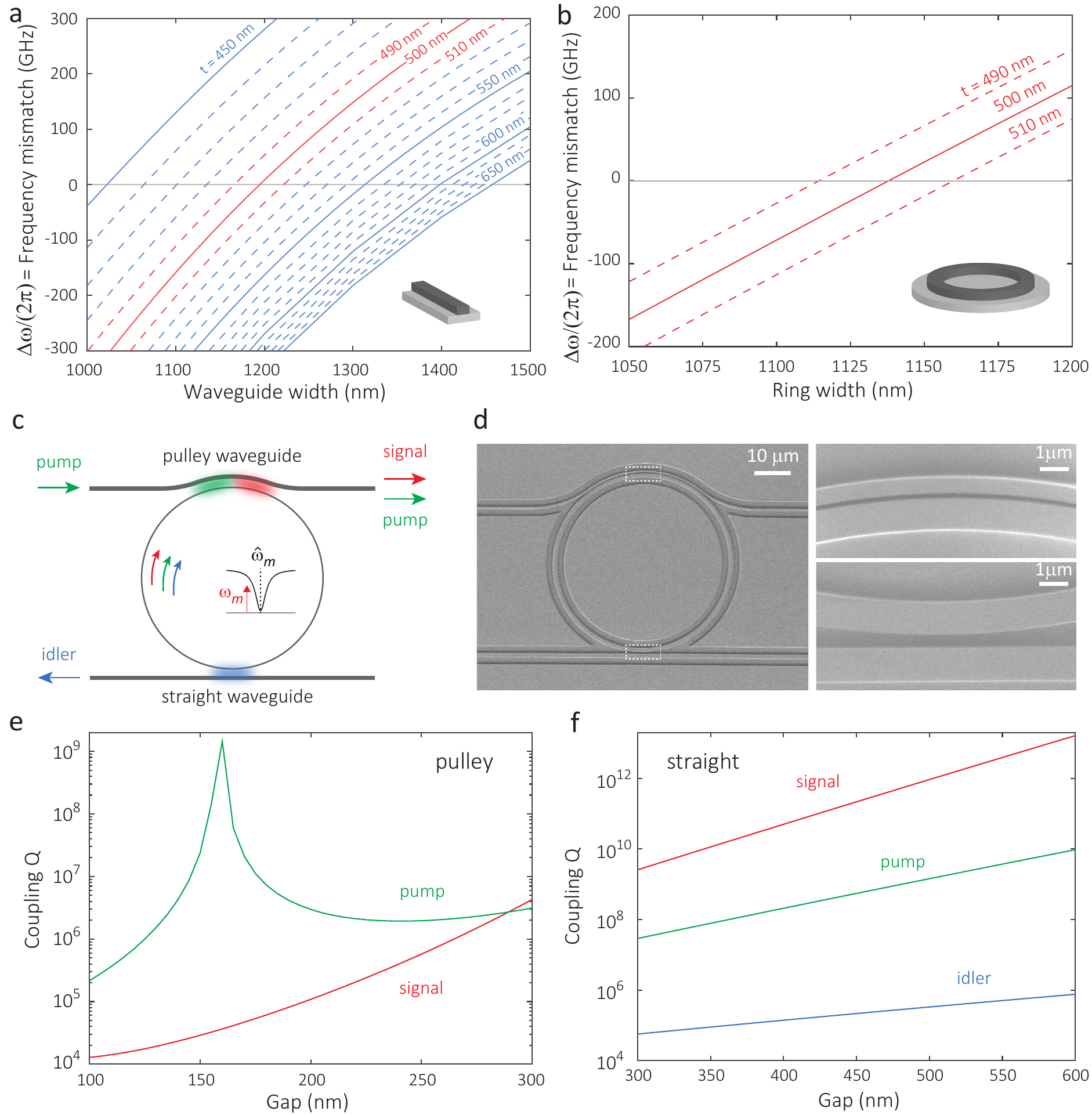}
\caption{\textbf{Design and Simulation}. \textbf{a, b}, Simulated frequency mismatch ($\Delta\omega = \omega_{p}-(\omega_{s}+\omega_{i})/2$) for phase-matched pump, signal, and idler modes ($2m_{p} = m_{s} + m_{i}$), as a function of ring waveguide width ($x$-axis) and thickness $t$ (different displayed curves).  Calculations are based on (a) a model in which straight waveguide dispersion (no bending effect) is used, and (b) a full model in which the ring dispersion is used.  The signal photon is at 668~nm, the idler photon is in the 1550~nm band, and the pump is in the 930~nm band. \textbf{c}, Waveguide-resonator coupling scheme. \textbf{d}, Scanning-electron-microscope images of a fabricated device with pulley waveguide on top, for injecting the 930~nm band pump and extracting the 660~nm band signal, and straight waveguide on bottom, for extracting the 1550~nm band idler. \textbf{e, f} Simulation of the coupling $Q$s of the pulley waveguide and straight waveguide, for the pump/signal and pump/signal/idler, respectively (the pulley waveguide does not support a mode at the idler wavelength).}
\label{fig:Fig2}
\end{center}
\end{figure*}
\end{center}

Most demonstrated microresonator-based SFWM devices have not required extensive dispersion engineering, largely because the frequency separation between the pump, signal, and idler has been relatively small (e.g., all lie within the telecommunications C-band)~\cite{caspani_integrated_2017}.  In such a scenario, mode number matching is trivially satisfied by signal and idler modes whose azimuthal orders are separated from the pump by an equal number, so that ${m_s-m_p = m_p - m_i}$.  Frequency matching of these modes is satisfied in a resonator with sufficiently low dispersion over the spectral window of interest, and does not present significant difficulty if the overall spectral window is small enough.

\begin{center}
\begin{figure*}
\begin{center}
\includegraphics[width=\linewidth]{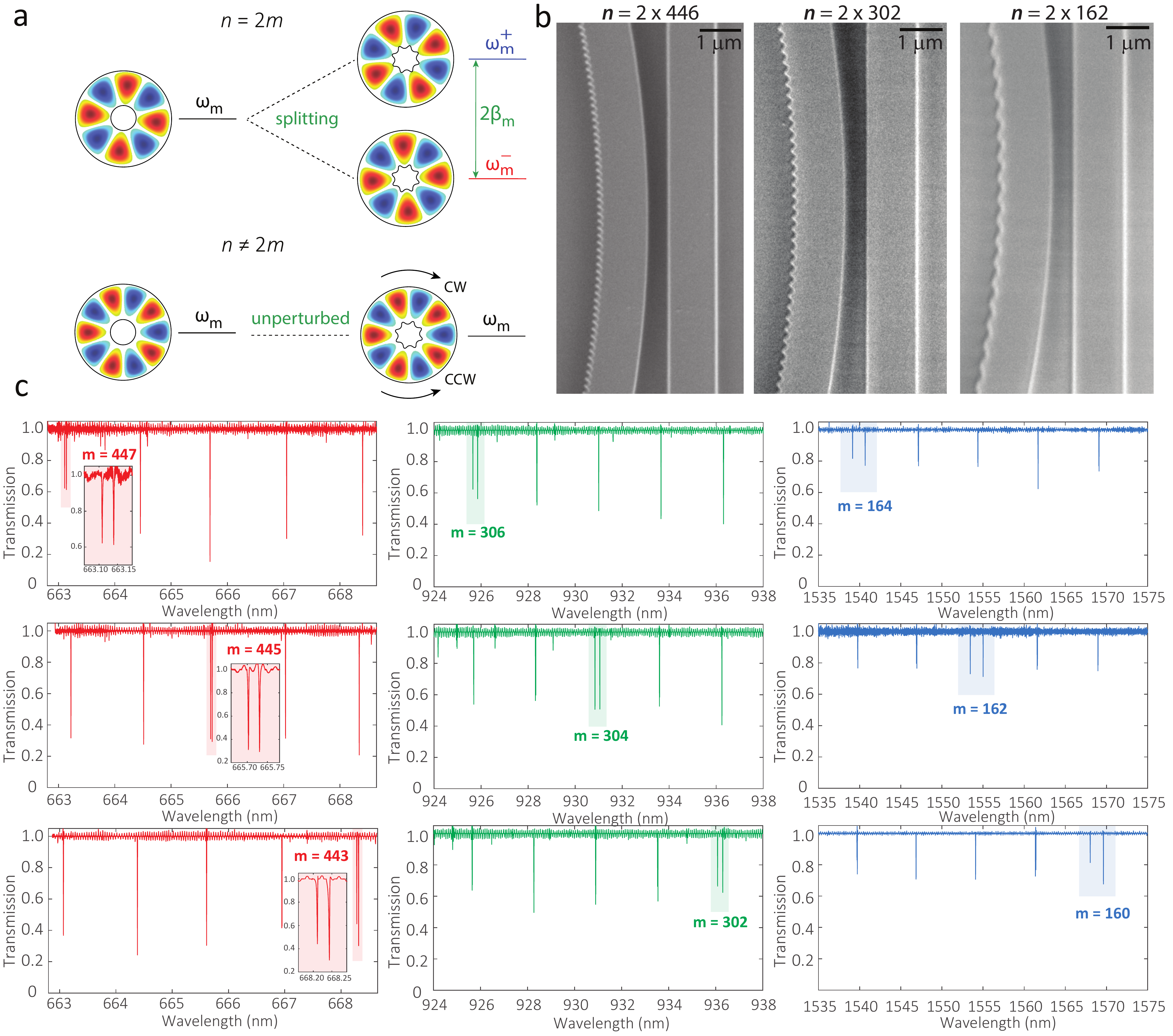}
\caption{\textbf{Mode Number Identification}. \textbf{a}, Principle of azimuthal mode number identification by mode splitting. The mode splitting is created by periodic modulation of the inside boundary of the microring, of the form $\delta r = \alpha {\rm{cos}}(n\phi)$.  When the modulation index $n$ is in phase with the azimuthal order of the mode $m$ (top), so that $n = 2m$, frequency splitting occurs.  When $n \neq 2m$, the optical modes are unperturbed (bottom). \textbf{b}, Zoomed-in SEM images of the mode splitting devices in the signal (left), pump (center), and telecom (right) bands, in which an inner modulation amplitude $\alpha = 50$~nm is applied.  The approximate modulation period in the three cases is 170~nm, 250~nm, and 460~nm, respectively. \textbf{c}, Measured transmission spectra from Si$_3$N$_4$ microrings fabricated with the inner boundary modulation, in the signal (visible), pump (930~nm), and telecom bands, where the modulation has been varied across the spectra to target the labeled mode numbers.}
\label{fig:Fig3}
\end{center}
\end{figure*}
\end{center}

In contrast, our SFWM process is distinguished by the large frequency separation between the optical modes, corresponding to over one hundred free spectral ranges (FSRs), which is greater than the scan range of any single laser. This makes counting of mode number offsets, to find the conditions for which ${m_s-m_p = m_p - m_i}$, very difficult. We instead identify the absolute azimuthal numbers of the targeted modes (${m_s}$, ${m_p}$, and ${ m_i}$), using a novel method described in the Mode Identification Section below.

\begin{center}
\begin{figure*}
\begin{center}
\includegraphics[width=\linewidth]{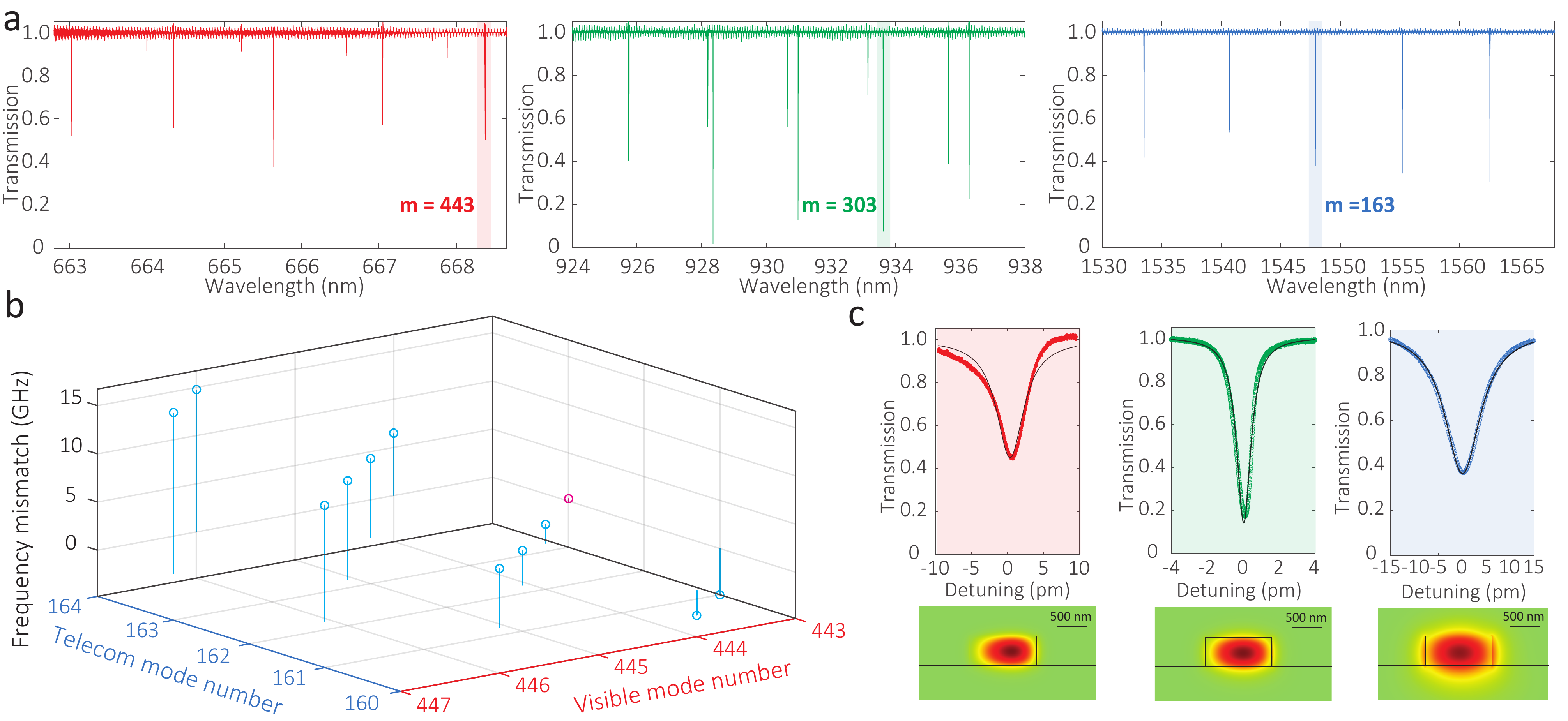}
\caption{\textbf{Phase/Frequency- Matching}. \textbf{a}, Transmission spectrum of the microring resonator in the signal (visible), pump (930~nm), and idler (telecom) bands. The targeted modes for photon pair generation are highlighted, with their azimuthal mode numbers labeled. \textbf{b}, Frequency mismatch $\Delta\omega = \omega_{p}-(\omega_{s}+\omega_{i})/2$, determined from wavemeter measurements in which the pump mode is fixed ($m_{p}$ =303) and the signal and idler modes vary in a way to maintain mode number matching ($2m_{p} = m_{s} + m_{i}$). \textbf{c}, Zoomed-in transmission spectra of the targeted optical modes, with Lorentzian fits in black. Insets show cross-sectional views of the simulated dominant electrical field component (in the radial direction) of each mode.}
\label{fig:Fig4}
\end{center}
\end{figure*}
\end{center}

The mode identification procedure, together with dispersion engineering to ensure that the targeted modes adequately satisfy frequency matching for the desired SFWM process (discussed below in the Dispersion Design Section), enables the generation of visible-telecom photon pairs within the ring resonator.  However, for the overall chip to be a bright and power-efficient photon pair generator, the coupling into and out of the resonator via access waveguides must also be optimized. As described in the Coupling Design Section, we utilize an add-drop geometry, together with the ability to cut off long wavelength modes in one of the waveguides as a result of its asymmetric cladding (top air cladding, bottom SiO$_2$ cladding) to separate the generated visible and telecom photons into different output waveguides.

\noindent \textbf{Dispersion Design}
The core of the device is a high-$Q$ Si$_3$N$_4$ microring with three key parameters - Si$_3$N$_4$ film thickness, ring width, and ring radius. The dispersion of the optical modes in the device is fully determined by these three parameters, given ellipsometric measurements of the wavelength-dependent refractive index of the Si$_3$N$_4$ and SiO$_2$ layers. We first carry out a waveguide simulation with the first two parameters to estimate the frequency mismatch, as shown in Fig.~\ref{fig:Fig2}(a). The frequency mismatch is calculated by the finite-element method as ${\Delta \omega = [(n_s \omega_s+n_i \omega_i)/ (2 n_p)] - \omega_p}$, where $n_s$, $n_p$, and $n_i$ are the effective indices at the frequencies $\omega_s$, $\omega_p$, and $\omega_i$ (corresponding wavelengths of  668.4 nm, 934.0 nm, and 1550.0 nm, respectively, for signal, pump, and idler modes). For a device with a larger thickness, the ring width needs to be smaller for the geometric dispersion to compensate the material dispersion. At a Si$_3$N$_4$ thickness of 500~nm, the targeted ring width is about 1200~nm. We then run full simulations for the Si$_3$N$_4$ microring with a radius of 25~${\rm \mu m}$ to find the optical modes around the targeted wavelengths, which have mode numbers of 443, 303, and 163, respectively. These mode numbers satisfy $2m_{p} = m_{s} + m_{i}$, and their corresponding resonances are close to the targeted frequencies (within one FSR). The frequency mismatches calculated from the ring simulations ${\Delta\omega = \omega_p - (\omega_s +\omega_i)/2}$ are plotted in Fig.~\ref{fig:Fig2}(b), where the bending effect yields additional geometric dispersion that offsets the targeted geometry from that determined from waveguide simulations in Fig.~\ref{fig:Fig2}(a). For a device thickness of 500 nm, the targeted ring width is around 1140 nm.

\noindent \textbf{Coupling Design}
The wide spectral separation of the SFWM process in this work also poses a challenge for the resonator-waveguide coupling design. To couple visible, pump, and telecom modes efficiently, we employ two waveguides that separate the coupling tasks with the scheme shown in Fig.~\ref{fig:Fig2}(c). A scanning-electron-microscope (SEM) image of the device is presented in Fig.~\ref{fig:Fig2}(d). The top pulley waveguide is wrapped around the microring (Fig.~\ref{fig:Fig2}(e)) to provide the extra interaction length needed to couple the pump and visible modes efficiently at the same time. This waveguide has a width of 560~nm, which supports single mode operation for the pump mode, and is cut-off for the telecom mode, a consequence of the asymmetric cladding (above a certain wavelength, not even a fundamental mode is guaranteed). Figure~\ref{fig:Fig2}(g) shows the calculated coupling $Q$ (using the coupled mode theory approach as in Refs.~\onlinecite{shah_hosseini_systematic_2010,li_efficient_2016}) of this pulley waveguide for the pump and signal modes, as a function of coupling gap and for a coupling length of 33~${\rm \mu}$m. For gaps between 200~nm to 250~nm, the pump mode is close to critically-coupled and the signal mode is slightly over-coupled for intrinsic $Q$s around one million. The bottom straight waveguide (Fig.~\ref{fig:Fig2}(f)) is used to out-couple the telecom mode and it has a width of 1120~nm. This width supports single mode operation at telecom wavelengths, and as shown in Fig.~\ref{fig:Fig2}(h), the coupling gap for the straight waveguide can be tuned to couple the telecom mode only, while not coupling pump and signal modes (they are severely undercoupled), for optical modes with intrinsic $Q$ around one million.

\noindent \textbf{Mode Identification}
The dispersion design presented in the previous sections is targeted to specific mode numbers and cavity frequencies.  Such unambiguous knowledge of optical mode numbers is absent in traditional transmission measurements of whispering gallery mode resonators. Here, we employ a method of mode splitting to identify the mode numbers~\cite{lu_selective_2014}, by introducing a sinusoidal modulation of the form $\delta r = \alpha {\rm{cos}}(n\phi)$ to the inside boundary of the microring (Fig.~\ref{fig:Fig3}(a)-(b)). When the modulation index $n$ does not line up with the azimuthal order $m$ of the optical mode (bottom panel of Fig.~\ref{fig:Fig3}(a), $n \neq 2m$), the perturbation is not in phase with the optical mode, and thus any reflection induced by the perturbation does not coherently build up. Therefore, in this case, the optical mode (both clockwise and counterclockwise) remains unperturbed. In contrast, when the modulation index lines up with the azimuthal order of the optical mode so that $n = 2m$, the phase degeneracy is removed and the clockwise and counterclockwise optical modes are renormalized into two modes (top panel of Fig.~\ref{fig:Fig3}(a)). One mode is located near the outside of the modulation and has a larger frequency, while the other mode is located near the inside of the modulation and has a smaller frequency. This splitting in frequency is proportional to the modulation amplitude. At a modulation amplitude of 50~nm (Fig.~\ref{fig:Fig3}(b)), the frequency splitting is much larger than the cavity linewidth and is therefore visible in the transmission spectra. Thus, by changing the modulation period, the optical mode numbers can be identified, as shown in Fig.~\ref{fig:Fig3}(c) for modes in the signal, pump, and idler bands. The $Q$ of modes in the ring-width-modulated resonators is still reasonably high ($Q>10^5$), so that one could envision using this approach to split modes that are some small number of FSRs away from the targeted ones, enabling their mode identification.  However, as we would like to use as high-$Q$ resonances as possible for pair generation, we instead use the mode splitting approach on test devices that are fabricated on the same chip as the devices in which we generate photon pairs.  The good reproducibility of device fabrication across a chip ensures that these test mode splitting devices have spectra that are well-matched to the photon pair generation devices.

\noindent\textbf{Phase/Frequency-Matching}
Through the mode number identification process, it is straightforward to find the targeted modes and calculate their frequency mismatch. Figure~\ref{fig:Fig4}(a) shows the normalized cavity transmission for a fabricated device in the signal, pump, and idler bands, respectively. The frequency mismatch of various configurations that satisfy mode number matching is shown in Fig.~\ref{fig:Fig4}(b). The targeted optical modes, shaded in Fig.~\ref{fig:Fig4}(a), have azimuthal mode numbers ${m_s=443}$, ${m_p=303}$, and ${m_i=163}$. The wavelengths of the modes, measured by a wavemeter with an accuracy of 0.1~pm, are 668.3789~nm, 933.6211~nm, and 1547.8960~nm. The loaded $Q$s of the three modes, displayed in Fig.~\ref{fig:Fig4}(c), are ${\rm (1.52~\pm~0.02) \times 10^5}$, ${\rm (1.04~\pm~0.02) \times 10^6}$, and ${\rm (1.93~\pm~0.04) \times 10^5}$, corresponding to loaded cavity linewidths of ${\rm (2.95~\pm~0.04)}$~GHz, ${\rm (0.31~\pm~0.01)}$~GHz, and ${\rm (1.00~\pm~0.02)}$~GHz, respectively. The uncertainties indicate ${\rm 95 ~ \%}$ confidence ranges of the nonlinear fitting. The frequency mismatch of the targeted modes ($\Delta\omega = \omega_p - (\omega_s +\omega_i)/2$) is only  ${\rm (0.16~\pm~0.04)}$~GHz (colored by red in Fig.~\ref{fig:Fig4}(b)), which is smaller than the cavity linewidths. Thus both phase- and frequency-matching are satisfied and the targeted SFWM process is permitted.

\begin{center}
\begin{figure*}
\begin{center}
\includegraphics[width=\linewidth]{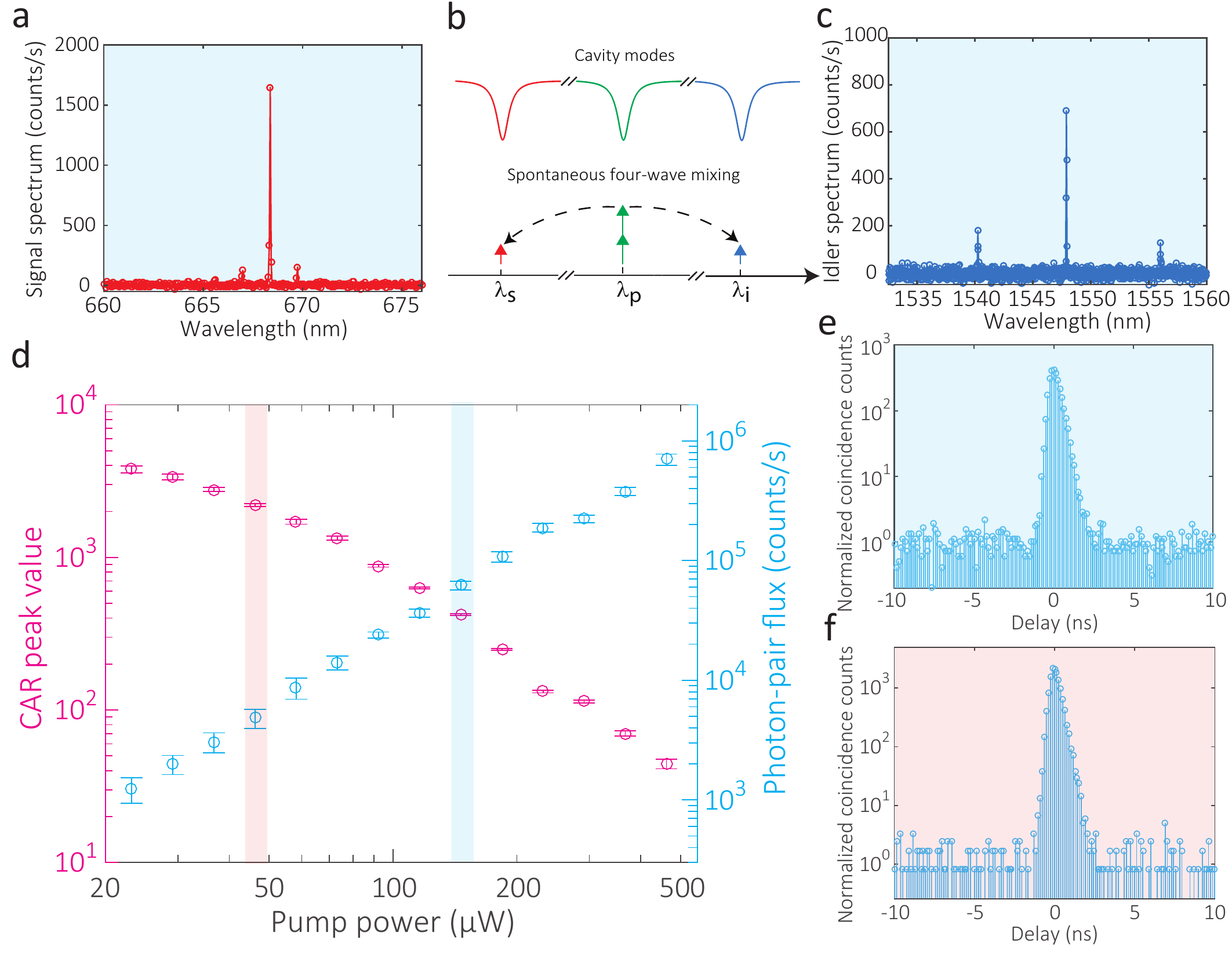}
\caption{\textbf{Visible-Telecom Photon Pair Generation}. \textbf{a, c,} Spectrum of spontaneous four-wave mixing in the visible and telecom bands, respectively, based on the photon pair generation scheme shown in \textbf{b}. \textbf{d,} Pump-power dependence of the visible-telecom photon pair source coincidence-to-accidental ratio (CAR) and on-chip photon pair flux.  Error bars are one standard deviation uncertainties resulting from fluctuations in the detected count rates. \textbf{e, f,} Coincidence counting trace for the two highlighted points in \textbf{d} (background color matches the highlighting color). The time bin is 128 ps. The mean background level (i.e., accidental counting) is normalized to 1 in these plots.}
\label{fig:Fig5}
\end{center}
\end{figure*}
\end{center}

\noindent\textbf{Photon Pair Generation}
Given the match of mode numbers and cavity frequencies as shown in Fig.~\ref{fig:Fig4}, we now consider the generation of photon pairs. Continuous-wave light from an external cavity diode laser is used to pump the cavity mode at 933.6~nm, and the spectrum of SFWM is measured using two grating spectrometers, one to cover the visible wavelength band and the other to cover the telecommunications wavelength band (see Supplementary Information).  Emission from cavity modes at 1547.9~nm and 668.4~nm is observed in the spectrometers (spectral resolution is $\approx$100~pm), along with two adjacent sets of pairs (Fig.~\ref{fig:Fig5}(a)-(c)). These auxiliary pairs are both mode number matched, but are weaker in amplitude due to the larger frequency mismatch. For example, the photon pair at 667.0 nm ($m_{s}$ = 444) and 1555.2 nm ($m_{i}$ = 162) has a frequency mismatch of 3.86 GHz, as shown in Fig.~\ref{fig:Fig4}(d), which is about 10${\times}$ larger than that of the brightest pair.

We next use coincidence counting measurements to characterize the quality of the generated photon pairs.  The large spectral separation between pump, signal, and idler makes this task straightforward, as the photons can be easily separated by edgepass filters or broadband demultiplexers.  Moreover, the spectra in the signal and idler bands exhibit limited influence from amplified spontaneous emission (ASE) and Raman noise, which are adjacent to the pump band and do not extend far enough spectrally to strongly influence the pairs. As a result, no broadband noise is observed in either the visible or telecom bands.

\begin{center}
\begin{figure*}
\begin{center}
\includegraphics[width=\linewidth]{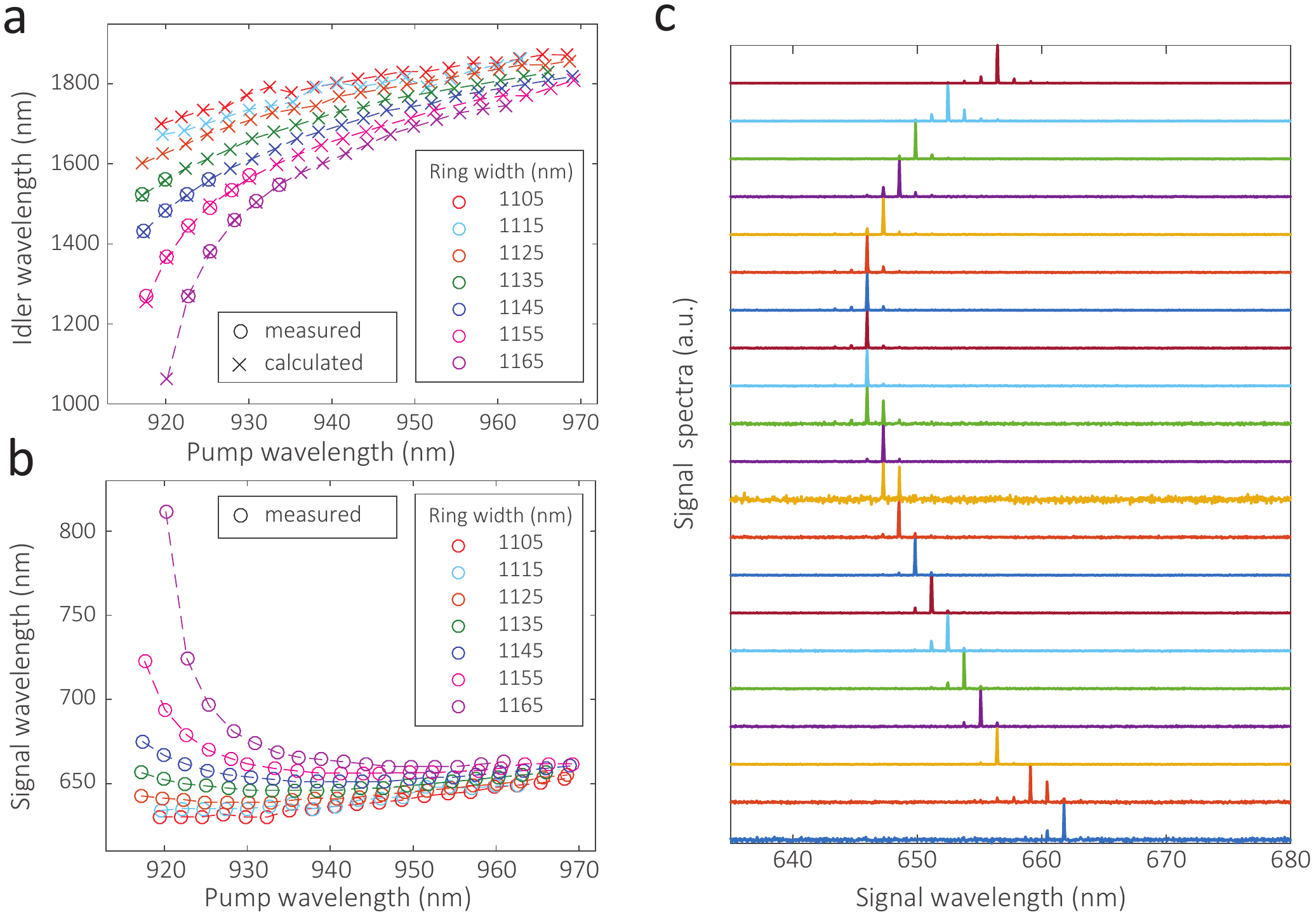}
\caption{\textbf{Wavelength Tuning of the Photon Pair Sources}. Dependence of \textbf{a} idler wavelength and \textbf{b} signal wavelength, generated by pumping microrings with varying ring widths on cavity modes across the 920~nm to 970~nm pump band. Measured data points are shown by $\circ$ symbols.  As the telecom band spectrometer is limited to the 1200~nm to 1600~nm range, idler wavelengths that lie above this wavelength range are calculated (shown by $\times$ symbols) based on the measured pump and signal wavelengths. \textbf{c}, Spectrum of SFWM in the visible wavelength band for a fixed ring width and a series of different pump wavelengths.}
\label{fig:Fig6}
\end{center}
\end{figure*}
\end{center}

For coincidence counting, two filters are used to select the targeted spectral channels before single photon detection (see Supplementary Information for details). Figure~\ref{fig:Fig5}(d) presents the pump-power-dependence of the photon-pair flux at the waveguide output and the coincidence-to-accidental ratio (CAR). When the pump power is around 46 $\rm \mu W$, the CAR value is about ${\rm (2200~\pm~50)}$ and the photon pair flux is about ${\rm (4.8~\pm~0.9) \times 10^3}$, where the uncertainties are one standard deviation values based on the fluctuation in detected count rates. The highest CAR measured is ${\rm (3800~\pm~200)}$ at a pair flux of ${\rm (1.2~\pm~0.3) \times 10^3}$ counts per second. This CAR value is two orders of magnitude higher than those of previously demonstrated visible-telecom photon pair sources based on photonic crystal fiber~\cite{clausen_source_2014} and millimeter-scale crystalline disks~\cite{schunk_interfacing_2015}. When the pump power is near 146 $\rm \mu W$, the CAR value is about ${\rm (422~\pm~5)}$ and the photon pair flux is about ${\rm (6.2~\pm~0.5) \times 10^4}$ counts per second.  At pump powers above $\rm 290~\mu W$, the photon pair flux shows a weaker increase than expected based on the slope of the lower power data, and is due to optical parametric oscillation in the pump band (see Supplementary Information).  This effectively serves as a competing process that results in a reduced conversion efficiency from the pump to visible-telecom photon pairs.  Finally, Fig.~\ref{fig:Fig5}(e)-(f) shows normalized coincidence spectra that correspond to two of the aforementioned cases, with CAR values of ${\rm (422~\pm~5)}$ and ${\rm (2200~\pm~50)}$, respectively.  The asymmetry in the coincidence peak is a result of the difference in photon lifetimes and detector timing jitters in the telecom and visible bands.

\noindent\textbf{Wavelength-Tunable Photon Pairs}

We have thus far focused on the generation of photon pairs according to a design that connects the 660~nm band with the 1550~nm band. However, the approach we have developed for engineering these sources is general, and one can easily envision extending it to connect other sets of wavelengths of interest.  As an example, we demonstrate tuning of the photon pair wavelengths through two experimental parameters, the pump wavelength (adjustable on a given device) and ring width (adjustable between devices on the same chip).  Figure~\ref{fig:Fig6} shows the results, in which the pump wavelength is varied between 920~nm and 970~nm, and the ring width between 1105~nm and 1165~nm.  As shown in Fig.~6(b)-(c), the visible photons shift from 630 nm to 810 nm, while the accompanying idler photon wavelength changes from 1800 nm to 1200 nm (Fig.~\ref{fig:Fig6}(a)). These sources span a significant fraction of the physical systems shown in Fig.~\ref{fig:Fig1}(c), including systems like the nitrogen vacancy center in diamond, the silicon vacancy center diamond, and the D1 and D2 transitions in Rb.

\noindent \textbf{Discussion}
We have demonstrated, for the first time, the ability to engineer on-chip nanophotonic resonators to create photon pair sources with a wide spectral separation between signal and idler, so that one photon is in the visible and the other in the telecom.  We use the Si$_3$N$_4$/SiO$_2$/Si platform to enable this work, due to its large Kerr nonlinearity, broadband optical transparency, and strong ability to engineer dispersion through geometric (waveguiding and bending) effects.  After performing electromagnetic simulations to identify sets of microring resonator modes that can satisfy both phase-matching (mode number matching) and frequency matching, we utilize a mode identification technique that allows us to experimentally determine absolute azimuthal mode numbers in fabricated devices, thereby paving the way for the experimental demonstration of photon pair generation.  Our photon pair sources generate bright and pure visible-telecom photon pairs with unprecedented CAR values at sub-mW pump power. By adjusting the pumping wavelength and device geometry, we show flexible tuning of the source wavelengths, with the visible photon of the pair moving from 630~nm to 810~nm. Through further development, we anticipate that these sources can be used in efforts to link distant quantum systems via entanglement swapping~\cite{halder_entangling_2007}.

\noindent \textbf{Methods}
\\
\noindent \textbf{Device Fabrication}
The device layout was done with the "Nanolithography Toolbox," a free software package developed by the NIST Center for Nanoscale Science and Technology~\cite{coimbatore_balram_nanolithography_2016}. The ${\rm Si_3N_4}$ layer is deposited by low-pressure chemical vapor deposition of a 500~nm thick ${\rm Si_3N_4}$ layer on top of a 3~${\rm \mu}$m thick ${\rm SiO_2}$  layer on a 100~mm Si wafer. The wavelength-dependent refractive index and the thickness of the layers are measured using a spectroscopic ellipsometer, with the data fit to an extended Sellmeier model. The device pattern is created in ZEP520A positive-tone resist by electron-beam lithography~\cite{NIST_disclaimer_note}. The pattern is then transferred to ${\rm Si_3N_4}$ by reactive ion etching using a ${\rm CF_4/CHF_3}$ chemistry. The device is chemically cleaned to remove deposited polymer and remnant resist, and then annealed at 1100~${\rm ^{\circ} C}$ in an ${\rm N_2}$ environment for 4 hours. An oxide lift-off process is performed so that the devices are with air-cladding on top but the input/output waveguides are with oxide-cladding on top. The edge of the chip is then polished for lensed-fiber coupling. After the polishing, the chip is annealed again at 1100~${\rm ^{\circ} C}$ in an ${\rm N_2}$ environment for 4 hours.

\noindent \textbf{CAR}
The CAR values in Figure 5(d) are calculated by CAR=(C-A)/A, where C and A are the overall and accidental coincidence counts extracted from the peak and background of the coincidence counting spectra (Fig. 5(e,f)), respectively. The one standard deviation uncertainty in CAR ${\rm \sigma_{CAR}}$ is given by ${\rm \sigma_{CAR}/CAR=\sqrt{(\sigma_C/C)^2+(\sigma_A/A)^2} \approx \sigma_A/A}$, where $\sigma_E$ and $\sigma_A$ are the one standard deviation uncertainties in the measured overall coincidence counts and accidental coincidence counts, respectively. The above approximation is made possible as ${\rm \sigma_C/C \gg \rm \sigma_A/A}$. A is the mean value taken over a 4~$\mu$s time window in the coincidence counting trace. ${\rm \sigma_A}$ is determined from 20 sections of coincidence counting backgrounds, each of which spans 200~ns.

\putbib

\noindent \textbf{Acknowledgements} X.L., Q.L., G.M., and A.S. acknowledge support under the Cooperative Research Agreement between the
University of Maryland and NIST-CNST, Award no. 70NANB10H193.

\noindent \textbf{Author Contributions} X.L. led the design, fabrication, and measurement, with assistance from Q.L., D.A.W., A.S., G.M., V.A., and K.S. X.L. and K.S. wrote the manuscript, and K.S. supervised the project.

\noindent \textbf{Additional Information} Correspondence and requests for materials should be addressed to X.L. and K.S.

\noindent \textbf{Competing financial interests} The authors declare no competing financial interests.

\end{bibunit}

\newpage
\onecolumngrid \bigskip

\begin{bibunit}

\begin{center} {{\bf \large Supplementary Information}}\end{center}

\setcounter{figure}{0}
\makeatletter
\renewcommand{\thefigure}{S\@arabic\c@figure}

\setcounter{equation}{0}
\makeatletter
\renewcommand{\theequation}{S\@arabic\c@equation}

\section{Comparison to related results}
The photon pair flux and the coincidence-to-accidental ratio (CAR) are the two main metrics by which we characterize the performance of our photon pair source. At high pump powers, the increase in pair flux is accompanied by a rise in accidental coincidence counts, limiting the achievable CAR values because of the increase in multi-pair generation inherent to spontaneous parametric processes, as well as the generation of noise photons or imperfect spectral rejection of residual pump light.  At low pump powers, the decrease in pair flux is accompanied by an almost complete suppression of multi-pair events, so that CAR values are primarily limited by detector dark counts and imperfect filtering.

To get a sense for how our photon pair source performance compares with that of related sources demonstrated in the literature, Figure S1(a) compares the CAR and the detected/generated pair flux of our photon source with the reported values from previous visible-telecom photon sources, which have primarily been created through spontaneous parametric downconversion in $\chi^{(2)}$ media.  To achieve the narrow bandwidth photons required for most quantum memories, various schemes have been adopted, including cavity-enhanced systems that incorporate a bulk quasi-phase-matched crystal~\cite{fekete_ultranarrow-band_2013,slattery_narrow-linewidth_2015,rielander_cavity_2016}, direct spectral filtering of broadband photons generated by a quasi-phase-matched waveguide or bulk crystal~\cite{clausen_source_2014}, and mm-scale whispering gallery mode resonators~\cite{schunk_interfacing_2015}. To gauge the performance of our pair source within the landscape of chip-integrated pair sources based on silicon nanophotonics, in Fig. S1(b), we consider several telecom-band photon pair sources that have been demonstrated.  We include data from two recent studies based on silicon nitride waveguides~\cite{zhang_correlated_2016} and hydex microrings~\cite{reimer_integrated_2014}, along with a number of different sources based on silicon microdisks/rings~\cite{clemmen_continuous_2009,azzini_classical_2012,engin_photon_2013,guo_impact_2014,harris_integrated_2014,gentry_quantum-correlated_2015, grassani_micrometer-scale_2015,jiang_silicon-chip_2015,wakabayashi_time-bin_2015,savanier_photon_2016,ma_silicon_2017}.

\begin{center}
\begin{figure*}
\begin{center}
\includegraphics[width=0.9\linewidth]{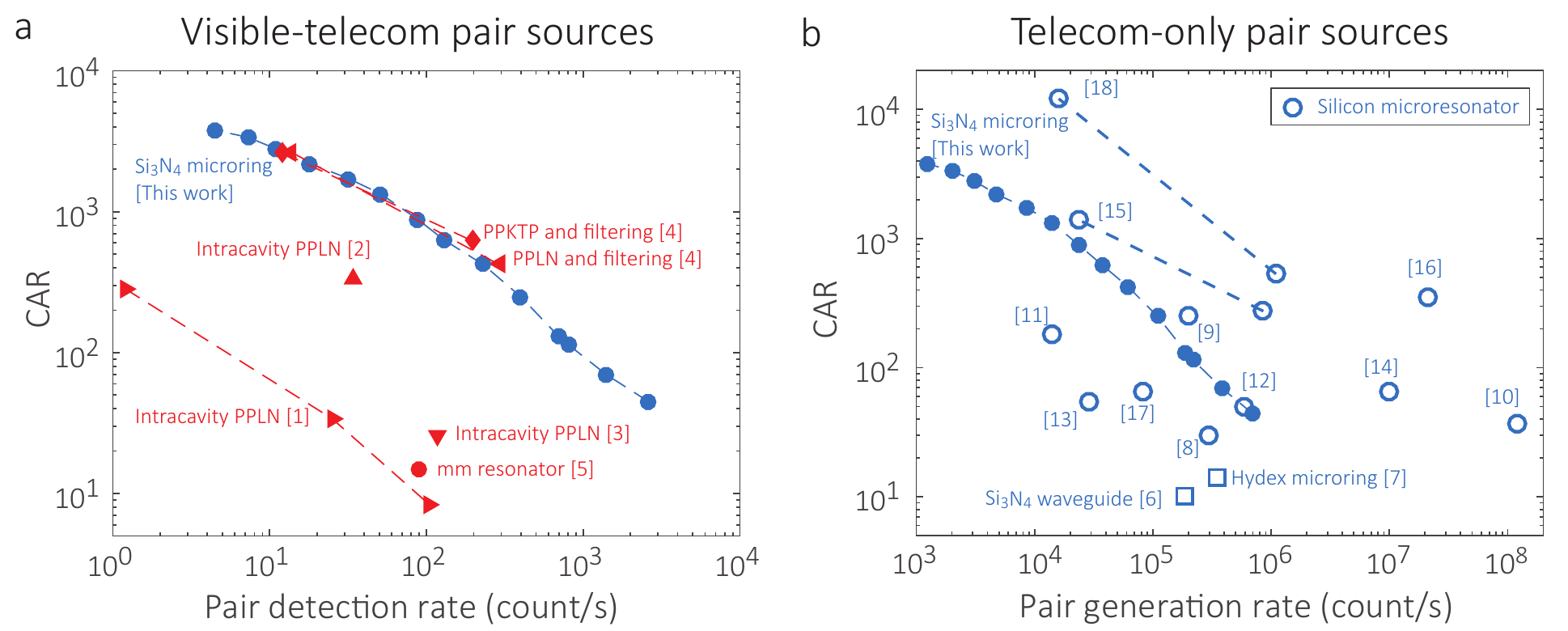}
\caption{\textbf{Photon pair source comparison}. A comparison of CAR and pair detection/generation rate of our source (blue solid circles) with \textbf{a}, reported values for other visible-telecom pair sources (red circles and triangles), and \textbf{b}, silicon (blue open circles) and Si$_3$N$_4$/hydex (blue open squares) telecom-only pair sources. Dashed lines are used to link the data points reported in the same work.}
\label{fig:FigS1}
\end{center}
\end{figure*}
\end{center}

We see that our source performs favorably against existing sources. In comparison with the different visible-telecom sources that have been demonstrated (Fig.~S1(a)), we achieve a combination of CAR and pair detection rate that are as good as has been previously documented in more mature, macroscopic (non-integrated) systems.  In comparison to chip-integrated nanophotonic devices, our source performance greatly improves upon previous results on telecom-only sources in silicon nitride and hydex, and begins to approach the best results reported for silicon-based, telecom-only sources~\cite{ma_silicon_2017}.


\section{Experimental setup}
The device is tested with the experimental setup shown in Figure S2, which illustrates the measurement configurations for cavity transmission, spectrum of spontaneous four-wave mixing, and coincidence counting. For the cavity transmission measurement, three tunable continuous-wave lasers are used to measure the transmission spectra of signal, pump, and idler bands, respectively. The laser wavelengths of cavity modes are calibrated by a wave meter with an accuracy of 0.1~pm. The pump laser is attenuated to sub-milliwatt levels with the polarization adjusted to transverse-electric-mode polarization. The pump light and the generated photon pairs are coupled on and off the chip by lensed optical fibers with a focus diameter of $\approx2.5~\mu$m. The fiber-chip coupling losses are 3.9~dB/2.3~dB/4.2~dB per facet for signal/pump/idler respectively. No filters are required before the device, since both the laser's amplified-spontaneous-emission noise and the Raman noise generated in the optical fibers have a frequency spectrum that is far from the visible and telecom bands. Signal and idler photons are coupled by separate waveguides and lensed fibers. The waveguides have photon extraction efficiencies of $\rm 84~\%/20~\%$ for signal/idler photons (this coupling level can be adjusted to trade off photon extraction efficiency for higher cavity $Q$ and hence narrower photon linewidth). The signal (visible) photons are separated from the residual pump by a dichroic mirror and a shortpass filter in free space. The idler (telecom) photons are separated by a longpass filter, which consists of two 980~nm/1550~nm fiber wavelength division multiplexers to yield a pump isolation over 90~dB. After these filtering stages, the spontaneous four-wave mixing spectra in the signal and idler bands is directed into a grating spectrometer and recorded by a liquid-nitrogen-cooled InGaAs photodiode array (telecom) and a TE-cooled silicon CCD (visible). For the coincidence counting measurement, tunable bandpass filters with bandwidths around 100 GHz are used to separate the photon pairs from their adjacent modes. The filtering losses for signal/idler are 5.8~dB/5.6~dB in total. The signal photons are measured by a single photon avalanche diode (SPAD) with multi-mode fiber coupling and the idler photons are measured by a superconducting nanowire single photon detector (SNSPD) with single-mode fiber coupling. The detection efficiencies of the SPAD/SNSPD are $\rm 59~\%/$\rm 67~\%, calibrated at the signal/idler wavelengths. A polarization controller (with a loss of 0.7~dB) is used to optimize the polarization for the SNSPD. Taking into account fiber-chip coupling, optical component loss, and detection efficiencies, the overall detection efficiencies for the on-chip signal/idler photons are 12.0~dB/12.2~dB. The Klyshko efficiencies are $\rm 3~\%/0.2~\%$ for the signal/idler photons.

\begin{figure}[htb]
  \begin{minipage}[c]{0.6\textwidth}
  \begin{center}
        \includegraphics[width=0.9\linewidth]{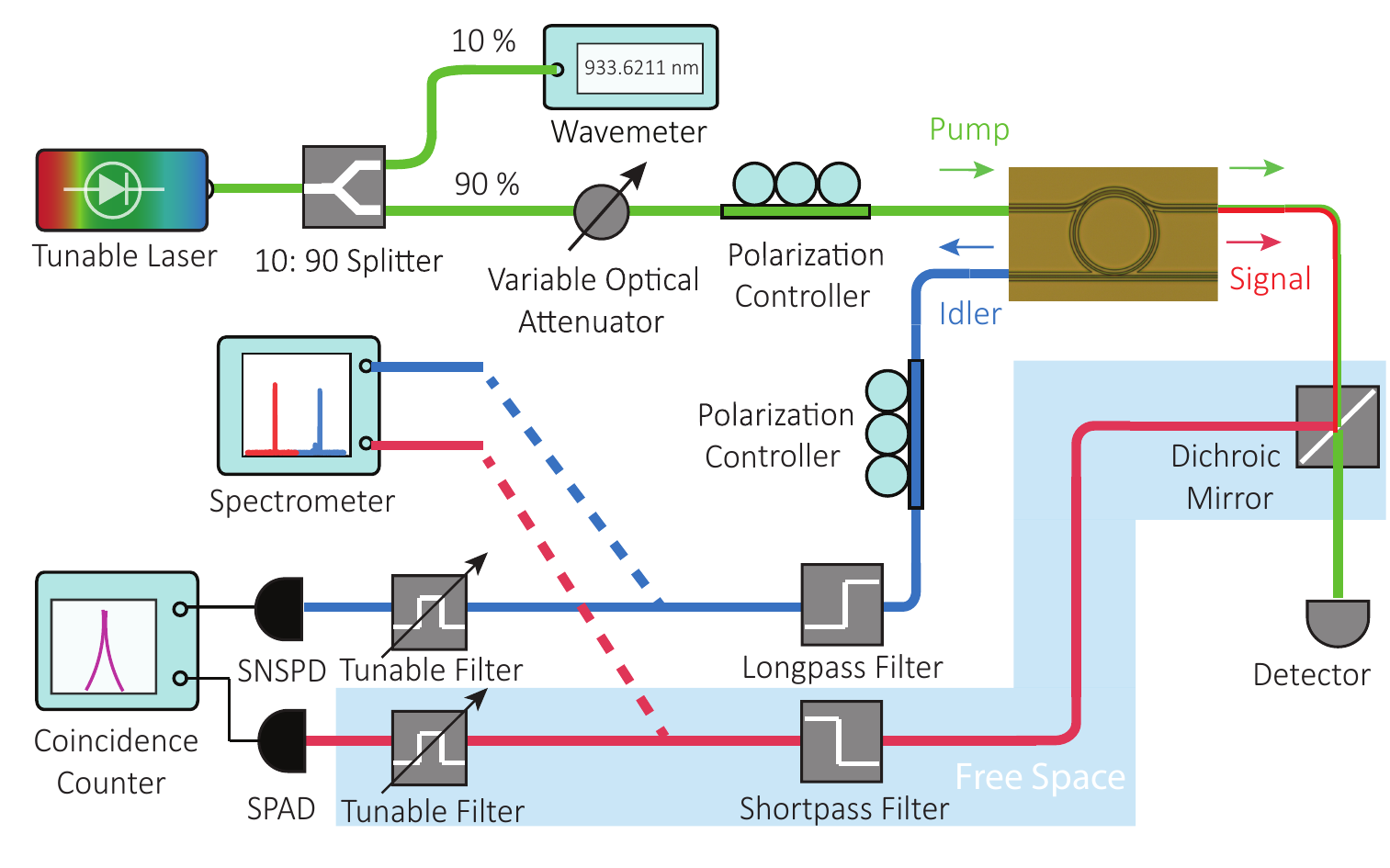}
  \end{center}
  \end{minipage}
  \begin{minipage}[c]{0.38\textwidth}
   \begin{center}
      \begin{tabular}{c|c c}
        \hline
                  & Signal & Idler \\ [0.5ex]
        \hline
        Fiber-chip coupling & 3.9 dB & 4.2 dB \\
        Polarization controller & - & 0.7 dB \\
        Filtering & 5.8 dB & 5.6 dB \\
        Detection & 59 \%/2.3 dB &  67 \%/1.7 dB \\
        \hline
        Channel loss & 12.0 dB & 12.2 dB \\
        Pair loss & \multicolumn{2}{c}{24.2 dB} \\
        \hline
      \end{tabular}
    \end{center}
  \end{minipage}
  \caption{\textbf{Experimental Setup and Itemized losses}. (Left) Figure of the experimental setup recording cavity transmission, SFWM spectra, and coincidence counting. Solid lines show the setup for the cavity transmission and the coincidence counting. Dashed lines indicate the setup for measuring the spectrum of spontaneous four-wave mixing. The shaded area indicates free-space operation, while the unshaded area indicates fiber optical components. SNSPD: superconducting nanowire single-photon detector. SPAD: single-photon avalanche photodiode. (Right) Table listing the itemized losses for signal and idler channels.}\label{Fig_S1}
\end{figure}

\section{Optical parametric oscillation}
In this section, we present experimental data indicating that optical parametric oscillation (OPO) occurs at the highest pump powers, helping to explain the decrease in photon-pair generation efficiency relative to the expectation at the high-power end of the plot in Figure 5(d). Figure S3(a) shows optical spectra around the pump band at various pump powers, recorded by an optical spectrum analyzer. When the pump power is $\gtrsim$~290 $\rm \mu W$ (yellow), parametric sidebands are generated with a spacing of 8 free spectral ranges (FSRs), and the power and number of sidebands increase with increasing pump power. At 460 $\rm \mu W$ power (blue), sidebands also emerge with a 1-FSR spacing alongside the 8-FSRs OPO modes.

As shown in Fig. S3(b), the pump cavity mode transmission contrast also exhibits significant variation with pump power. At the lowest pump powers, e.g., $\approx$15~$\mu$W, the cavity transmission shows the typical Lorentzian lineshape. As the optical pump power is increased, the cavity transmission exhibits the triangle shape that is characteristic of the bistability associated with the thermo-optic effect when increasing the pump laser wavelength. For powers below 290 $\rm \mu W$, the cavity transmission contrast is fixed, and the microring is slightly under-coupled with an on-resonance transmission value of 28~$\%$. When the optical power is further increased, parametric oscillation is initiated, and some fraction of the pump power is converted to parametric sidebands. This frequency conversion is effectively an additional loss channel for the pump, so that the cavity becomes more undercoupled (the total loss rate increases).  As a result, the cavity transmission contrast decreases and the on-resonance value increases to 32~\%, 37~\%, and 41~\% for the yellow, red, and blue traces, respectively. Thus, when the pump power increases beyond the parametric oscillation threshold, pump depletion due to this competing process becomes significant, resulting in a lower generation efficiency for the visible/telecom photon pairs.

\begin{center}
\begin{figure*}
\begin{center}
\includegraphics[width=\linewidth]{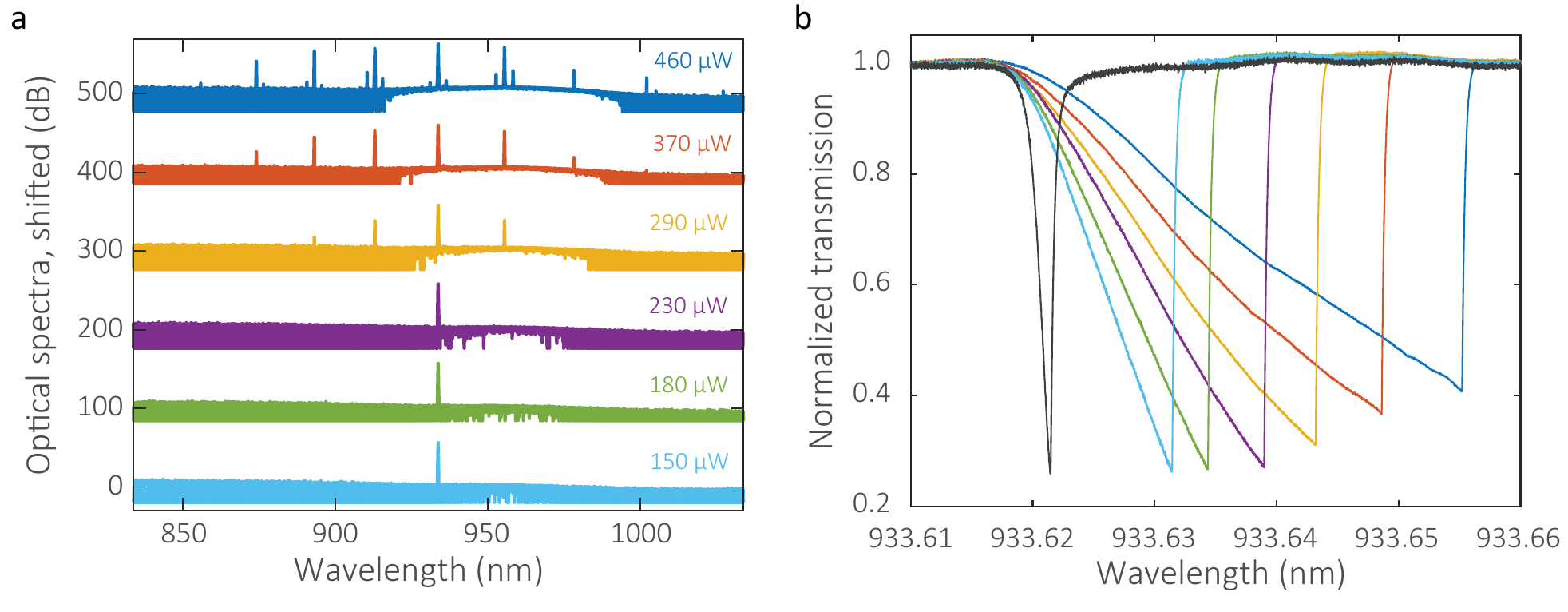}
\caption{\textbf{Optical Parametric Oscillation}. \textbf{a}, Optical spectra with pump power varied from 460 $\rm \mu W$ (top) to 150 $\rm \mu W$ (bottom) with a step of 1 dB. The spectra are spaced by 100 dB. \textbf{b}, Cavity transmission traces of the pump resonance at various pump powers. The black trace has a low pump power of 15 $\rm \mu W$ and exhibits no thermal bistability. The other traces have pump powers from 150 $\rm \mu W$ (left) to 460 $\rm \mu W$ (right).}
\label{fig:FigS3}
\end{center}
\end{figure*}
\end{center}

\putbib

\end{bibunit}

\begin{thebibliography}{10}
\bibitem{NIST_disclaimer_note}
The identification of any commercial product or trade name is used to foster
  understanding. Such identification does not imply recommendation or
  endorsement or by the National Institute of Standards and Technology, nor
  does it imply that the materials or equipment identified are necessarily the
  best available for the purpose.
\end{thebibliography}
\end{document}